\documentclass{article}
\usepackage{spconf,amsmath,graphicx}
\usepackage{cite}
\usepackage{amsmath,amssymb,amsfonts}
\usepackage{algorithmic}
\usepackage{graphicx}
\usepackage{textcomp}
\usepackage{dsfont}
\usepackage{epstopdf}
\usepackage{subcaption}
\usepackage{tikz}
\usepackage{color}
\usepackage{hyperref}
\usepackage{cleveref}

\title{Sketched RT3D: How to reconstruct billions of photons per second}
\name{Juli\'an Tachella, Michael P. Sheehan, Mike E. Davies\sthanks{This work was supported by the ERC Advanced grant, project C-SENSE, (ERC-ADG-2015-694888). Mike E. Davies is also supported by a Royal Society Wolfson Research Merit Award. Code is available at \url{https://gitlab.com/Tachella/real-time-sp-lidar}.
© 2022 IEEE.  Personal use of this material is permitted.  Permission from IEEE must be obtained for all other uses, in any current or future media, including reprinting/republishing this material for advertising or promotional purposes, creating new collective works, for resale or redistribution to servers or lists, or reuse of any copyrighted component of this work in other works}%
  }
	\address{School of Engineering\\
	University of Edinburgh\\
	Edinburgh, UK}

\DeclareMathOperator*{\argmin}{arg\,min}
\newcommand{\RNum}[1]{\uppercase\expandafter{\romannumeral #1\relax}}
\DeclareMathOperator{\sinc}{sinc}

\def\refs{{\boldsymbol{\alpha}}}
\def\btheta{{\boldsymbol{\theta}}}
\def\deps{{\boldsymbol{t}}}

\def\z{{\mathbf z}}

\def\E{{\mathbb E}}

\begin{document}
	%
	\maketitle
	
\begin{abstract}
Single-photon light detection and ranging (lidar) captures depth and intensity information of a 3D scene. Reconstructing a scene from observed photons is a challenging task due to spurious detections associated with background illumination sources. To tackle this problem, there is a plethora of 3D reconstruction algorithms which exploit spatial regularity of natural scenes to provide stable reconstructions. However, most existing algorithms have computational and memory complexity proportional to the number of recorded photons. This complexity hinders their real-time deployment  on modern lidar arrays which acquire billions of photons per second. Leveraging a recent lidar sketching framework, we show that it is possible to modify existing reconstruction algorithms such that they only require a small sketch of the photon information. In particular, we propose a sketched version of a recent state-of-the-art algorithm which uses point cloud denoisers to provide spatially regularized reconstructions. A series of experiments performed on real lidar datasets demonstrates a significant reduction of execution time and memory requirements, while achieving the same reconstruction performance than in the full data case.
\end{abstract}

\section{Introduction}
\begin{figure}[h!]
	\centering
	\includegraphics[width=.5\textwidth]{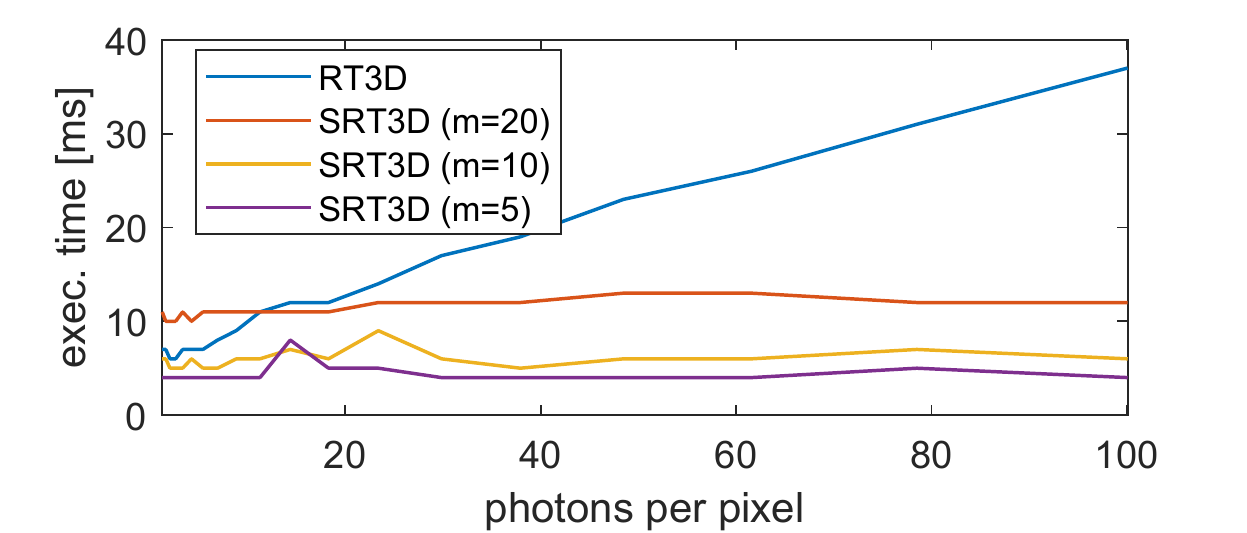}
	\caption{Execution time of  RT3D~\cite{tachella2019rt3d} and the proposed sketched SRT3D as a function of the mean number of photons per pixel. RT3D suffers from a linear complexity, whereas SRT3D only depends on the number of sketches $m$. }
	\label{FIG:timings}
\end{figure}

Single-photon lidar technology enables multiple important applications, ranging from autonomous driving \cite{hecht2018lidar,rapp2019spm} to tropical archaeology~\cite{canuto2018maya}. This sensing modality can provide long-range information~\cite{Pawlikowska:17} with millimetre precision~\cite{McCarthy:09} while using eye-safe laser power levels. Depth information is obtained by measuring the round-trip time-of-arrival (ToA) of laser pulses using a time-correlated single-photon counting system. Recovering 3D information from the photon detections can be very challenging in scenarios where the number of collected photons associated with the signal of interest is very low or in the presence of strong ambient illumination which generates spurious  detections. Leveraging the spatial regularity of natural scenes, several algorithms\cite{shin2015photon,altmann2016lidar,altmann2016target,shin2016computational,rappfew,Lindell:2018:3D,tachella2018manipop} have been proposed to provide stable 3D reconstructions in these settings.  Existing reconstruction algorithms require access to either the ToA of all the detected photons or a ToA histogram. Both memory requirements and computational complexity scale at least linearly with these parameters~\cite{tachella2019rt3d} (see~\cref{FIG:timings}). This dependency hinders their use in modern lidar arrays, which record an ever-increasing number of photons per second~\cite{RoyoLidarReview}.

\begin{figure*}[t!]
	\centering
	\includegraphics[width=\textwidth]{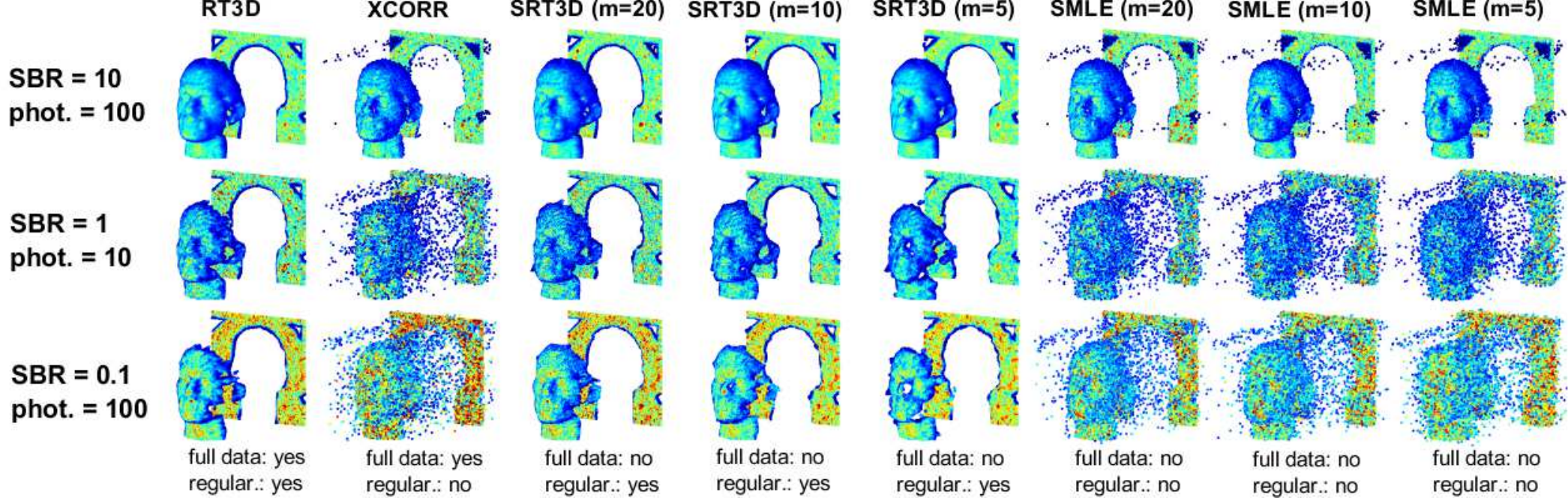}
	\caption{3D reconstructions obtained by the proposed sketched RT3D algorithm for different sketch sizes $m$ and other competing algorithms. The proposed SRT3D method incorporates spatial regularization, providing stable reconstructions in settings with low SBR or low number of measured photons. }
	\label{FIG:all heads}
\end{figure*}
\par A simple solution to the computational and memory bottlenecks consists of aggregating the fine-resolution timing data into histograms with a small number of bins, at the cost of sacrificing depth resolution~\cite{henderson2018imager}. This approach results in a trade-off between compression and temporal resolution, leading to suboptimal reconstructions which do not make full use of the high resolution potential of the lidar device. In contrast, a novel sketching based approach was recently proposed in~\cite{sheehan2021sketchedlidar} as a  solution to the data transfer bottleneck that does not suffer from an inherent trade-off between compression and temporal resolution. A compact representation of the ToA data is computed using the detected photon time-stamps in an online manner for each individual pixel. The size of this representation scales with the parameters of the ToA model (i.e. the positions and intensities of the objects) and is independent of both the temporal resolution  and the number of photons. However, the reconstruction methods for sketched data presented in~\cite{sheehan2021sketchedlidar} do not incorporate any spatial regularization (pixels are processed independently), and thus can provide unstable reconstruction when the signal-to-background ratio (SBR) is low (see~\cref{FIG:all heads}).

\par In this paper, we extend the sketched lidar framework to spatially regularized reconstruction algorithms. In particular, we present a novel sketched-based real-time algorithm
whose complexity is agnostic of the number of recorded photons. The proposed algorithm paves the way for real-time 3D reconstruction of complex scenes for any number of recorded photons. The main contributions of the paper are:
\begin{itemize}
    \item We propose a novel approach to regularized 3D reconstruction with memory and computation requirements that are independent of the total number of observed photons.
    \item We evaluate the robustness of the sketched reconstruction algorithm using real lidar datasets, demonstrating that the proposed approach achieves the same reconstruction performance as in the full data case.
\end{itemize}
\noindent The paper is organised as follows. In \Cref{Sec: Sketched Lidar}, we introduce the sketched lidar framework. In \Cref{sec:sketch reg}, we present a novel reconstruction algorithm with spatial regularization which only relies on sketched data. In \Cref{sec: results}, we analyse the performance of the sketched algorithm on synthetic and real datasets. Conclusions are discussed in \Cref{sec: conclusion}.

\section{Pixelwise Sketched Lidar}\label{Sec: Sketched Lidar}
An individual single-photon lidar pixel is associated with $n$ time-stamps indicating the ToA of individual photons, which are denoted by $x_p$ for $1\leq p\leq n$. Assuming that there are $K$ distinct reflecting surfaces in the field of view of the pixel, we let $\deps = [t_1,\dots,t_{K}]^{\top}$ be a vector containing the depths of the surfaces,  $\refs = [\alpha_1,\dots,\alpha_{K}]^{\top}$ where $\alpha_k$ is the probability that the detected photon from the $k$th surface. We denote by $\alpha_0$ the probability that the detected photon is due to the background illumination. The ToA of the $p$th photon detected follows a mixture distribution~\cite{altmann2}
\begin{equation}
    \label{Eqn: Observ model}
    \pi(x_p| \refs, \deps)= \sum^K_{k=1}\alpha_k\pi_s(x_p|t_k)+\alpha_0\pi_b(x_p),
\end{equation} 
where $\sum^K_{k=0}\alpha_k=1$. The distribution of the photons originating from the signal is defined by $\pi_s(x_p|t)=h(x_p-t)/H$, where the impulse response of the system and its associated integral are denoted by $h$ and $H=\sum_{t=1}^T h(t)$, respectively. The distribution of photons originating from background sources is in general uniformly distributed, $\pi_b(x_p)=1/T$ over the interval $[0,T-1]$  (more complex models can also be accommodated). The parameters in \cref{Eqn: Observ model} are summarized by the tuple $\theta = 
(\deps,\refs,\alpha_0)$.

\par  Timing information is generally represented either as a list of $n$ ToA values~\cite{tachella2018manipop} or aggregated into a ToA histogram with $T'\leq T$ temporal bins~\cite{henderson2018imager}. Modern lidar devices acquire hundreds or thousands of photons $n$ in a short time frame, which hinders the use of ToA lists as the memory requirement scales linearly with the $n$. Histograms can alleviate the memory requirement by aggregating photon detections into coarse temporal bins, i.e.
\begin{equation}
    y_\ell = \sum_{p=1}^{n} \mathds{1}_{x_p \in [t_\ell, t_\ell + \Delta t]}
\end{equation}
for $\ell =1,\dots,T'$.
However, this strategy sacrifices depth resolution, leading to suboptimal reconstructions which do not exploit the full potential of the device.

\par In order to alleviate the trade-off between memory requirement and depth resolution, \cite{sheehan2021sketchedlidar} have recently introduced a novel representation of the timing information, whose size is independent of the number of acquired photons and does not incur in a significant loss of depth resolution.  
The compact representation, or so-called sketch, is computed as
\begin{equation}
    \label{Eqn: The sketch}
    z_\ell = \frac{1}{n}\sum^{n}_{p=1} e^{ {\rm i}\omega_{\ell}x_p},
\end{equation}
for $\ell=1,\dots,m$ where ${\rm i}=\sqrt{-1}$ and $m$ is the number of 
sketches. 
As with the coarse histogram, the sketch has the favourable property that it can be updated in an online fashion with each incoming photon throughout the duration of the acquisition time. Thereafter, only the resultant sketch $\z = [z_1,\dots,z_\ell]^{\top} \in \mathbb{C}^{m}$ needs to be stored and/or transferred off-chip to further estimate the parameters $\theta$ of the observation model.
\par The sketch is equivalent to the empirical characteristic function sampled at frequencies $\omega_{\ell}$, where $\Psi_\theta(\omega)=\E_{x\sim\pi}e^{ {\rm i}\omega x}$ is the corresponding expected characteristic function (CF) \cite{carrasco2000generalization}. The CF has the special property that it exists for all probability distributions and captures all the information of the distribution, providing a one-to-one correspondence. The CF of the observation model in (\ref{Eqn: Observ model}) is given by
\begin{equation}
\label{Eqn: Char function Obs Model}
\Psi_{\theta}(\omega)= \sum^{K}_{k=1}\alpha_k\hat{h}(\omega)e^{{\rm i}\omega t}+\alpha_0\sinc(\omega T/2),
\end{equation}
where $\hat{h}$ denotes the (discrete) Fourier transform of the impulse response function $h$.
It is well documented in the empirical characteristic function (ECF) literature e.g. \cite{gemmhall}, that a sketch $\z$ computed over a finite dataset $\mathcal{X}=\{x_1,\dots,x_n\}$, satisfies the central limit theorem. Formally, the sketch $\z$ converges asymptotically to a Gaussian random variable
\begin{equation}
\label{Eqn: sketch central limit theorem}
    \z\xrightarrow[]{\text{dist}}\mathcal{N}\big(\Psi_{\theta},n^{-1}\Sigma_{\theta}\big),
\end{equation}
where $\Sigma_{\theta}\in\mathbb{C}^{m\times m}$ depends on $\Psi_{\theta}(\omega)$.
Thus, the sketched lidar inference task reduces to solving the following optimization problem
\begin{equation}
    \label{Eqn: CL loss function}
 \argmin_{\theta} n\lVert\z -\Psi_{\theta}\rVert^2_\mathbf{W},
\end{equation}
where $\mathbf{W}\in\mathbb{C}^{m\times m}$ is a positive definite Hermitian weighting matrix, which can be chosen as the identity matrix for reduced computational load or the precision matrix $\mathbf{W}=\Sigma_\theta^{-1}$ for higher statistical efficiency. 
 \par The sampling scheme proposed in \cite{sheehan2021sketchedlidar} chooses the  frequencies $\omega_\ell=2\pi \ell/T$ for $\ell\in[1,T-1]$, such that $\Psi_\theta$ in \cref{Eqn: Char function Obs Model} is only sampled at regions where $\sinc(\omega T/2)=0$, and the resulting  sketch is effectively blind to background noise. 
 Fundamentally, the size of the sketch $m$ scales with the degree of parameters in the observation model (i.e. the number of surfaces in the scene) and, crucially, is independent of both the depth resolution and the number of photons $n$. Sketching therefore enables significant compression of the ToA data without sacrificing temporal resolution or estimation accuracy. In the next section, we show how spatial regularization can also be incorporated to the sketching framework.

\section{Multipixel sketched lidar} \label{sec:sketch reg}
Single-photon lidar devices acquire an array of $N_r\times N_c$ pixels.  We encompass all the parameters in a scene into $\boldsymbol{\theta} = (\theta_{1,1},\dots,\theta_{N_r,N_c})$. Due to the spatial regularity of natural scenes, parameters in neighbouring pixels are generally strongly correlated. This prior knowledge is exploited by several 3D reconstruction algorithms to improve the quality with respect to simple pixelwise depth estimation. Most algorithms solve the following optimization problem~\cite{rapp2019spm}
\begin{equation}
    \label{Eqn: typical}
      \argmin_{\btheta} \sum_{i,j}^{N_r,N_c} f_{\boldsymbol{y}_{i,j}}(\theta_{i,j}) + \rho(\btheta)
\end{equation}
where $\boldsymbol{y}_{i,j}$ denotes the observed histogram at the $(i,j)$th pixel, $f_{\boldsymbol{y}_{i,j}}(\theta_{i,j})$  are per-pixel data fidelity terms (negative log-likelihood of the ToA list or histogram observation models), and $\rho(\btheta)$ is a spatial regularization term which encodes the prior information about the spatial regularity of typical scenes. There has been significant efforts dedicated to the design of powerful regularizations $\rho(\btheta)$. The RT3D algorithm~\cite{tachella2019rt3d} exploits the plug-and-play framework~\cite{venkat2013pnp} together with a fast computer-graphics point cloud denoiser to design a regularizer that can capture the geometry of complex scenes with $K>1$ surfaces per pixel, while achieving real-time performance. However, existing algorithms (including RT3D) require multiple evaluations of the data fidelity terms, and thus suffer from large memory requirements and a computational complexity which is at least linear in the number of photon detections or histogram bins~\cite{tachella2019rt3d}.

Here we propose to replace the histogram-based loss in~\cref{Eqn: typical} for the more compact sketch loss in~\cref{Eqn: CL loss function}, while leveraging the spatial regularization penalty of existing methods. The proposed objective can be expressed as
\begin{equation}
    \label{Eqn: reg CL loss function}
      \argmin_{\btheta} \sum_{i,j}^{N_r,N_c} n_{i,j}\lVert\z_{i,j} -\Psi_{\theta_{i,j}}\rVert_{\mathbf{W}_{i,j}}^2 + \rho(\btheta)
\end{equation}
where $\z_{i,j}$ is the sketch associated with the $(i,j)$th pixel. The number of detected photons $n_{i,j}$ controls the trade-off between the data-fidelity and regularization terms. As the number of detected photons increases, the data fidelity term dominates~\cref{Eqn: reg CL loss function}, which tends to the  non-regularized problem in~\cref{Eqn: CL loss function}. In order to perform real-time reconstruction with an arbitrary number of photon detections, we propose a sketched version of the RT3D algorithm, which we name SRT3D. The proposed algorithm replaces the histogram-based likelihood of RT3D for the sketched loss of~\cref{Eqn: reg CL loss function} with $\textbf{W}_{i,j}$ set as the identity matrix for reduced computational load. 

\section{Experiments} \label{sec: results}
We evaluate the proposed SRT3D algorithm on two real datasets: a polystyrene head at a distance of 40 metres~\cite{altmann2016lidar} and a scene with two people walking behind a camouflage net at a distance of 320 metres~\cite{tachella2019rt3d}. We compare the proposed method with 3 other algorithms: traditional cross-correlation~\cite{McCarthy:09} (XCORR), pixelwise sketched reconstruction~\cite{sheehan2021sketchedlidar} (SMLE) which doesn't exploit spatial regularization (also with $\mathbf{W}$ set as the identity), and  RT3D which accesses the full fine-resolution ToA data. All the experiments were performed using an NVIDIA RTX 3070 laptop GPU. 

\begin{figure}[h!]
	\centering
	\includegraphics[width=.48\textwidth]{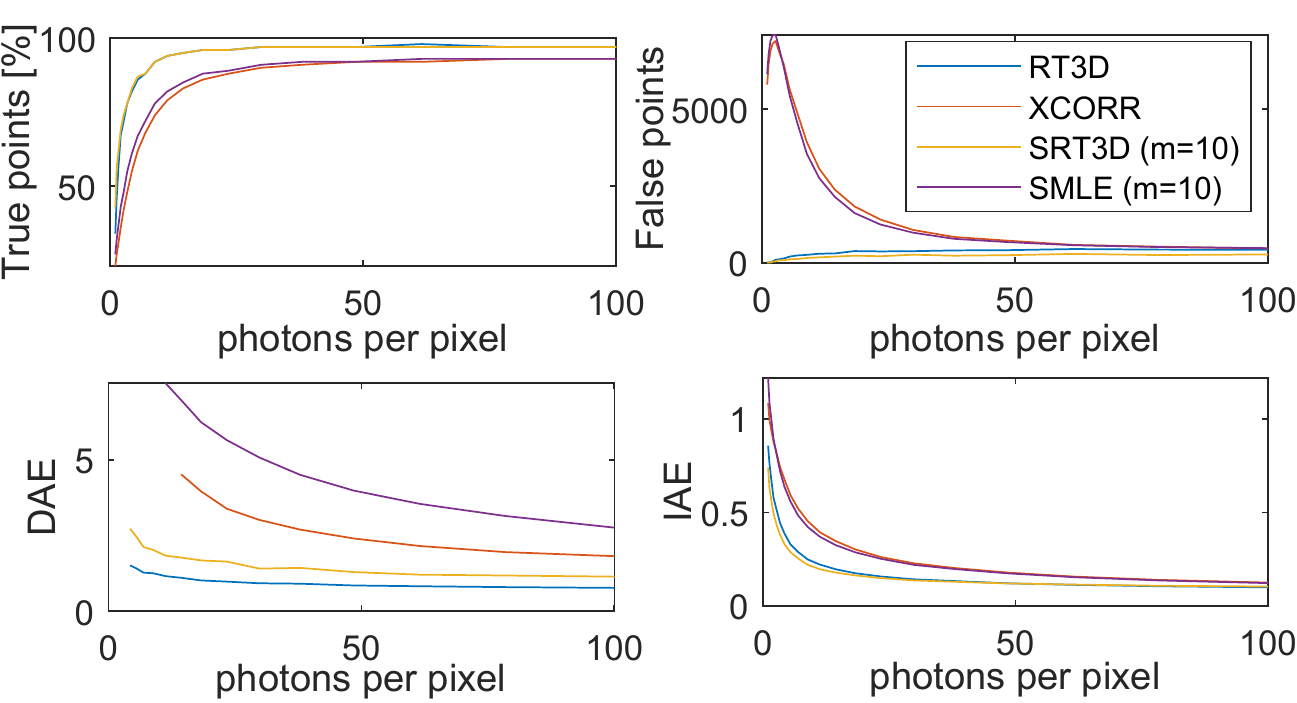}
	\caption{Performance of the evaluated algorithms for the polystyrene head dataset with SBR=1. }
	\label{FIG:metrics}
\end{figure}

The polystyrene head dataset has size of $141\times 141$ pixels with $T=4613$. Most of the pixels in this scene contain exactly $K=1$ surface. A ground-truth reference was obtained using using the standard cross-correlation algorithm on the raw ToA information (with high number of photons per pixel and high SBR). Using this reference and the observation model in~\cref{Eqn: Observ model}, we synthesized multiple datasets for different mean photons per pixel $n$ and SBR levels. \Cref{FIG:all heads} shows the 3D reconstructions obtained for SBR levels of 10, 1 and 0.1. All methods perform similarly when the number of photons and SBR are large. Interestingly, only $m=5$ sketches are enough to provide good reconstructions. However, when the scene contains a low number of photons or low SBR, pixelwise methods fail to provide good reconstructions, whereas both RT3D and SRT3D provide good reconstructions. In this challenging setting, a sketch of size $m=10$ sufficiently provides a reconstruction that has the same quality as the ones obtained in the full data case.
True and false detections, depth absolute error (DAE), and normalised intensity absolute error (IAE) (as defined in~\cite{tachella2019rt3d}) are presented in~\cref{FIG:metrics} for an SBR of 1 and different number of photons per pixel.

\Cref{FIG:timings} shows the execution time of RT3D and SRT3D as a function of the mean number of photons per pixel in the polystyrene head datasets. The GPU memory requirements of RT3D become prohibitive  if the number of observed photons is in the order of hundreds per pixel per frame, whereas the sketched version has a complexity which is agnostic of the number of photons, and can handle any number of photons in real-time. \Cref{tab:pixel timing} shows the execution time of SRT3D for increasing array sizes,  demonstrating that the proposed method can process up to $705\times 705$ arrays at 14 frames per second on a laptop computer. The datasets were generated by upsampling the head reference before the synthesis of photon detections.


\begin{table}[h]
\centering
\begin{tabular}{|l|c|c|c|c|c|}
\hline
$m$/pixels& \multicolumn{1}{l|}{$141^2$} & \multicolumn{1}{l|}{$282^2$} & \multicolumn{1}{l|}{$423^2$} & \multicolumn{1}{l|}{$564^2$} & \multicolumn{1}{l|}{$705^2$} \\ \hline
$m=5$ & 6 & 12 & 28 & 55 & 68 \\ \hline
$m=10$ & 7 & 18 & 35 & 60 & 88 \\ \hline
\end{tabular}
\caption{Execution time in milliseconds for different scene sizes in pixels obtained by the proposed sketched RT3D algorithm for sketch sizes $m$ of 5 and 10.}
\label{tab:pixel timing}
\end{table}

\begin{figure}[h!]
	\centering
	\includegraphics[width=.5\textwidth]{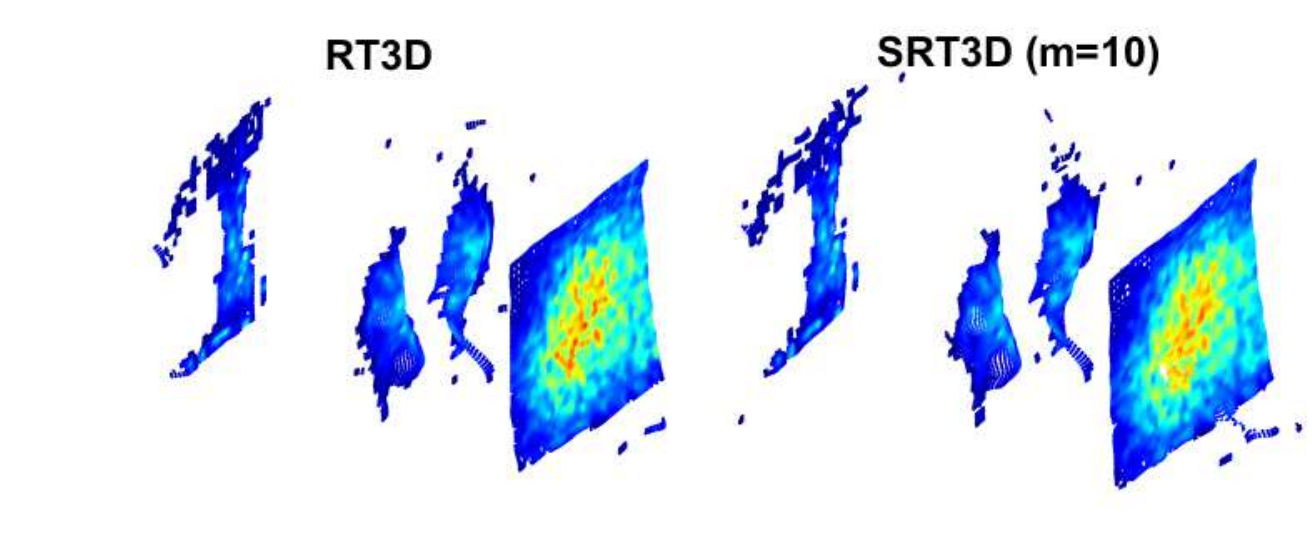}
	\caption{Reconstruction of the scene in~\cite{tachella2019rt3d} with 2 surfaces per pixel by the original RT3D and its sketched version. Using only $m=10$ sketches is enough to provide the same reconstruction quality.}
	\label{FIG:camou}
\end{figure}

\Cref{FIG:camou} shows the reconstructions by SRT3D and RT3D of one time frame of the camouflage dataset in~\cite{tachella2019rt3d}, which is composed of $32\times32$ pixels with $T=153$. Most of the pixels in the scene contain $K=2$ surfaces, which makes the reconstruction task more challenging. However, $m=10$ sketches are again sufficient to provide the same reconstruction quality as using the full $153$ bins. Although the original fine resolution $T$ is not large, the execution time of SRT3D with $m=10$ was 12 ms, whereas for RT3D it was 20 ms.

\section{Conclusions and future work} \label{sec: conclusion}
We have extended the sketched lidar framework to incorporate spatial regularization, demonstrating stable reconstructions in the presence of strong ambient illumination. While our results focus on a sketched version of the RT3D algorithm, the ideas presented here can be used to develop sketched versions of other existing regularized methods.

\section*{Acknowledgements}
We would like to thank the single-photon group of Heriot-Watt University for the head and camouflage data. 

\bibliographystyle{IEEEtran}
\bibliography{bibliography}

\end{document}